\documentclass[aps,twocolumn,superscriptaddress,showpacs]{revtex4}



\usepackage{natbib}

\usepackage{graphicx}
\usepackage{dcolumn}
\begin{document}


\title{Evolution of worldwide stock markets, correlation structure and correlation based graphs}

\author{Dong-Ming Song}
\affiliation{School of Business, East China University of Science and Technology, Shanghai 200237, China}
\affiliation{School of Science, East China University of Science and Technology, Shanghai 200237, China}

\author{Michele Tumminello}
\affiliation{Department of Social and Decision Sciences, Carnegie Mellon University, Pittsburgh PA 15213, USA}
\affiliation{Dipartimento di Fisica, Universit\`a di Palermo, Viale delle Scienze, Ed. 18, I-90128, Palermo, Italy}

\author{Wei-Xing Zhou}
\affiliation{School of Business, East China University of Science and Technology, Shanghai 200237, China}
\affiliation{School of Science, East China University of Science and Technology, Shanghai 200237, China}
\affiliation{Research Center for Econophysics, East China University of Science and Technology, Shanghai 200237, China}

\author{Rosario N. Mantegna}
\affiliation{Dipartimento di Fisica, Universit\`a di Palermo, Viale delle Scienze, Ed. 18, I-90128, Palermo, Italy}

\date{\today}

\begin{abstract}
We investigate the daily correlation present among market indices of stock exchanges located all over the world in the time period Jan 1996 - Jul 2009. We discover that the correlation among market indices presents both a fast and a slow dynamics. The slow dynamics reflects the development and consolidation of globalization. The fast dynamics is associated with critical events that originate in a specific country or region of the world and rapidly affect the global system. We provide evidence that the short term timescale of correlation among market indices is less than 3 trading months (about 60 trading days). The average values of the non diagonal elements of the correlation matrix, correlation based graphs and the spectral properties of the largest eigenvalues and eigenvectors of the correlation matrix are carrying information about the fast and slow dynamics of correlation of market indices. We introduce a measure of mutual information based on link co-occurrence in networks, in order to detect the fast dynamics of successive changes of correlation based graphs in a quantitative way.
\end{abstract}



\maketitle


\section{Introduction}

The  correlation structure of financial asset returns is informative for stock return time series \cite{Mantegna-1999-EPJB} (for a recent review see \cite{Tumminello-Lillo-Mantegna-2010-JEBO}), market index returns of stock exchanges located worldwide \cite{Bonanno-Vandewalle-Mantegna-2000-PRE,Maslov-2001-PA,Coelh-Gilmore-Lucey-Richmond-Hutzler-2007-PA,Gilmore-Lucey-Boscia-2008-PA,Eryigit-Eryigit-2009-PA} and currency exchange rates \cite{McDonald-Suleman-Williams-Howison-Johnson-2005-PRE,Gorski-Drozdz-Kwapien-2008-APPA}. The correlation based clustering procedures  \cite{Tumminello-Lillo-Mantegna-2010-JEBO} allow also to associate  correlation based networks with the correlation matrix. Useful examples of correlation based networks are the minimum spanning tree \cite{Mantegna-1999-EPJB}, graphs obtained by using thresholding procedures \cite{Onnela-Chakraborti-Kaski-Kertesz-2002-EPJB,Onnela-Chakraborti-Kaski-Kertesz-2003-PA} and the Planar Maximally Filtered Graph (PMFG) \cite{Tumminello-Aste-DiMatteo-Mantegna-2005-PNAS}.

In this paper we investigate the daily correlation present among indices of stock exchanges located all over the world. The study is performed by using the index time series of 57 different stock markets monitored during the time period Jan 1996 - Jul 2009. By investigating this set of stock market indices we discover that the correlation among world indices has both short term and long term dynamics. The long term dynamics is a slow monotonic growth associated with the development and consolidation of globalization while the short term dynamics is associated with events originated in a specific part of the world and rapidly affecting the entire system. Examples are the 1997 Asian crisis, the 1998 Russian crisis, the 2007 development of the subprime crisis and the onset of the 2008 global financial crisis. The presence of both short term and long term timescales in the dynamics of the correlations among world indices make difficult their analysis. In fact an estimation of the empirical correlation matrix minimizing the unavoidable statistical uncertainty associated with the evaluation needs a large number of records to be used in the time evaluation period. However, an extended time evaluation period reduces the ability to resolve the fast dynamics of correlation. In the present study, we first perform our analyses by using different time evaluation periods and we then analyze the dynamics of the correlation at the shortest time scale accessible by ensuring that the correlation matrix is invertible (the correlation matrix is no more invertible when the number of records in the time evaluation period is less then the number of elements in the investigated set). We provide empirical evidence that the short timescale of correlation among world indices can be less than 3 trading months (about 60 trading days) and that there are quite stable factors driving the dynamics of stock market indices located in specific regions of the world.

We also show that the interrelation between stock market indices can be efficiently described by using correlation based networks and principal component analysis. Unsupervised cluster detection is performed on a correlation based network obtained by using the correlation matrix estimated using all daily records available. The cluster detection is done by applying a community detection algorithm to the correlation based network. We show that the characteristics of fast dynamics of the interrelations among stock indices are well described by the PMFGs and by the two largest eigenvalues and eigenvectors of the correlation matrix. Abrupt short term alterations are detected at the onset of several financial crisis but the changes detected in the structure of graphs and in the principal component analysis profile are of difficult economic interpretation due to the high level of statistical uncertainty associated with the correlation estimation and because different events might be crisis specific and therefore specific only to each single event. To quantify in an efficient way successive changes of correlation based networks estimated with the shortest evaluation time period we introduce a new way to compute a mutual information measure between two networks based on link co-occurrence.

The paper is organized as follows. In Section \ref{S1:Data}, we briefly present the set of investigated data and we discuss the time scales of the dynamics of correlations of market indices. In Section \ref{S1:CorrGraph}, we analyze the unconditional correlation based graph associated with index returns and we perform a community detection on it. In Section \ref{S1:Events}, we discuss the short term dynamics of correlations of stock indices. In Section \ref{S1:MutualInfo}, we discuss the time evolution of PMFGs computed with the shortest evaluation time period and we compare successive networks by using a newly introduced mutual information measure based on link overlap. In Section \ref{S1:Spectral}, we investigate the dynamics of the largest eigenvalues and eigenvectors associated with correlation matrices computed at the shortest timescale. In the last Section we present our conclusions.

\section{Data and time scales}
\label{S1:Data}

In this study, we investigate a set of 57 stock market indices of 57 different exchanges located in several continents. The complete list of stock market indices is given in the appendix. Data are sampled daily.  We have selected these 57 stock market indices because we have access for them to a long time period ranging from January 1996 to July 2009. We perform our analysis on the daily logarithmic return, which, for each index $i$, is defined as:
\begin{equation}
  r_i(t)=\ln P_i(t)-\ln P_i(t-1),
\end{equation}
where $P_i(t)$ is the price of index $i$ on day $t$. Starting from the return time series we compute the correlation matrix of this multivariate set of data at time $T$ by using past return records sampled during evaluation time periods of different length $\Delta{T}$ ranging from 3 calendar months ($\Delta T=0.25$, approximately 60 trading days) up to 5 calendar years ($\Delta{T}=5$, approximately 1250 trading days). For each month $t$ (converted to $T$ in unit of years) and for each different evaluation time interval $\Delta T$, we compute the Pearson correlation coefficient
\begin{equation}
  c_{i,j}(T,\Delta T)=\frac{\langle[r_i(k)-\mu_i][r_j(k)-\mu_j]\rangle}{\sigma_i \sigma_j},
\end{equation}
where $\mu_i$ and $\mu_j$ are the sample means and $\sigma_i$ and $\sigma_j$ are the standard deviations of the two stock index time series $i$ and $j$ respectively.

\begin{figure}[htb]
  \includegraphics[scale=0.5]{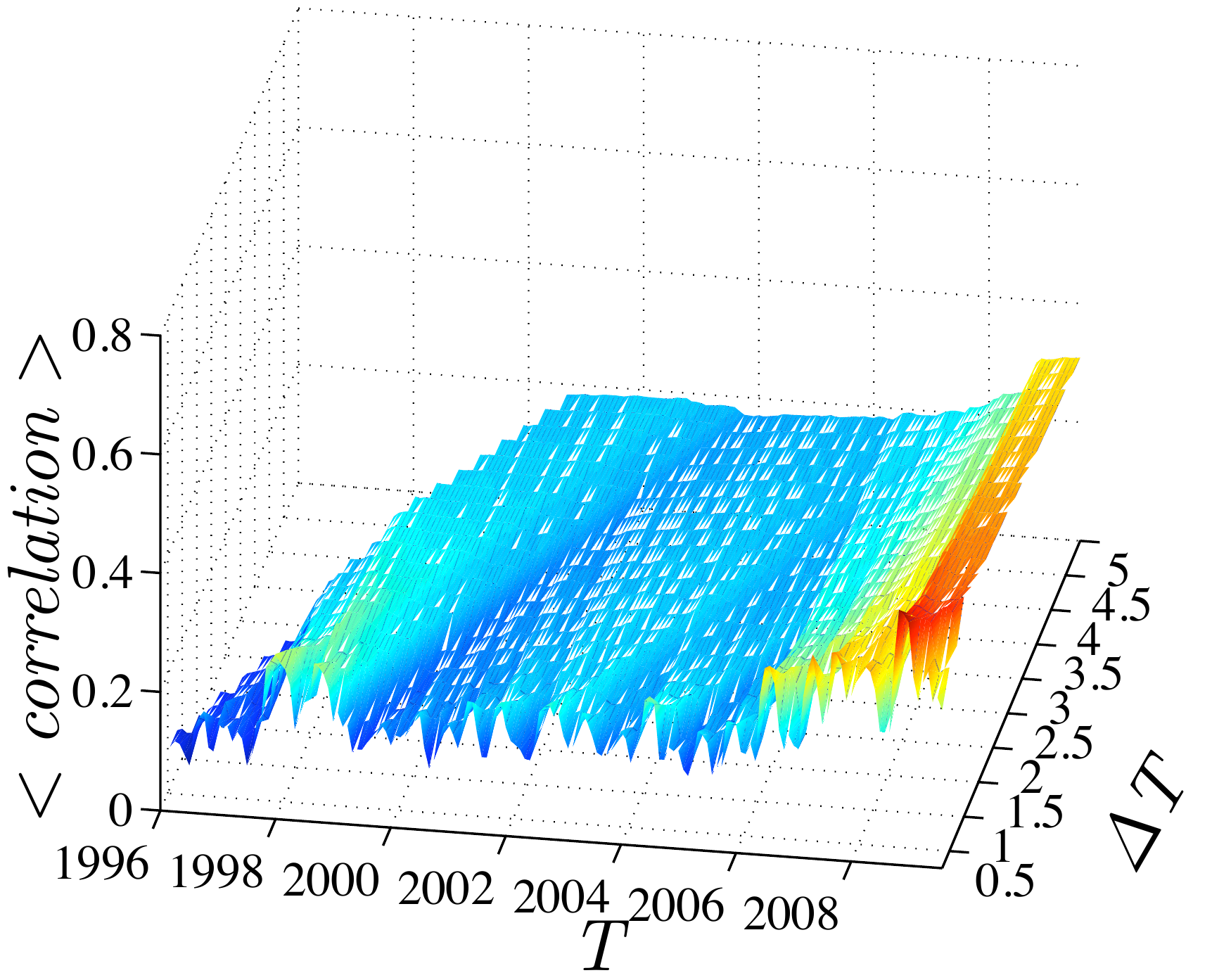}
  \caption{Average correlation of the non diagonal elements of the correlation matrix as a function of the evaluation time $T$ and of the evaluation time period $\Delta T$.}
  \label{fig1}
\end{figure}

\begin{figure}[htb]
  \includegraphics[scale=0.45]{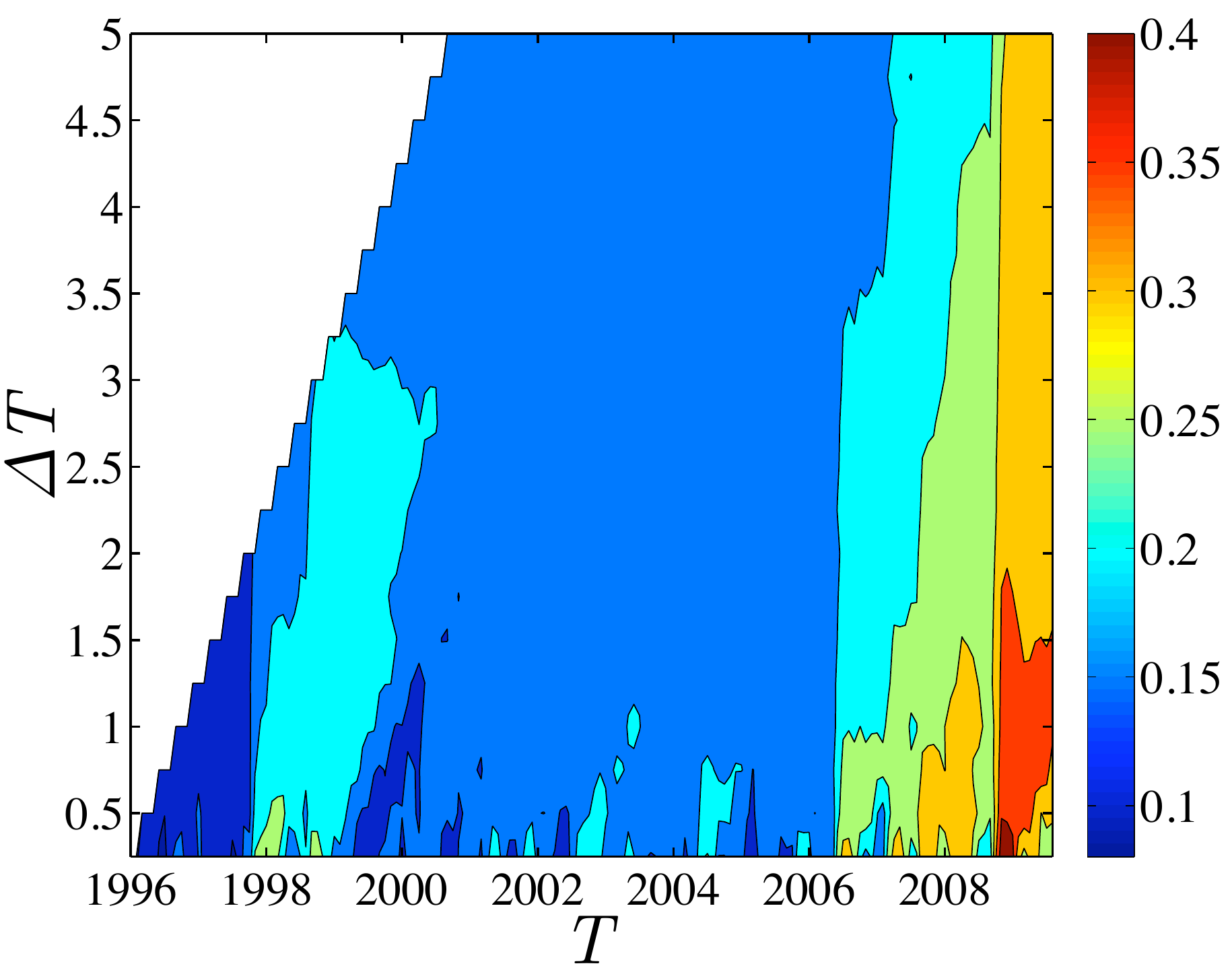}
  \caption{Contour plot of the average correlation of the non diagonal elements of the correlation matrix as a function of the evaluation time $T$ and of the evaluation time period $\Delta T$. The white region is the region where the past records are not enough to estimate the correlation matrix with the same statistical accuracy of other $T$ values for the same $\Delta T$.}
  \label{fig2}
\end{figure}

In Fig.~\ref{fig1} we plot the average correlation value $\langle{c_{i,j}(T,\Delta T)}\rangle$ of the non diagonal elements of the correlation matrix computed at time $T$ using a set of past daily records spanning a $\Delta T$ interval. In Fig.~\ref{fig2} we show the contour plot of the average correlation value of the correlation matrix. The results summarized in Figs. \ref{fig1} and \ref{fig2} shows that a dynamics is present in the time evolution of the correlation among the indices of different stock exchanges. Important aspects to be investigated concern both the fast and the slow dynamics of the correlations. Ideally one would like to estimate correlation among indices by using a short estimation interval. Unfortunately by using a short estimation interval the level of the statistical uncertainty in the estimation is increased and eventually one ends up with not well-characterized correlation matrices when the number of time records used in the estimation are less or close to the number of investigated market indices \cite{Ledoit-Wolf-2004-JMA}. On the other hand, when a long estimation interval is used successive estimations of the correlation are not independent and therefore localized jumps of the average correlation are smeared out over a long time period.

In fact, by looking at Figs.~\ref{fig1} and \ref{fig2}, we notice that both a short term and a long term dynamics is present in the evolution of correlation. We also conclude that to detect properly the short time scale of correlation dynamics we need to use a short evaluation time period because the contour plot of Fig. \ref{fig2} shows that the localization of the onset of big sizable changes are affected by the length of the evaluation time period $\Delta T$. For example the onset of the Asian 1997 crisis, the Russian 1998 crisis, the 2007 subprime crisis and the 2008 global crisis are quite clearly detected when an evaluation time period of 3 months is used whereas the onset is smeared out and postponed when longer evaluation time periods are used.

\section{Correlation based graphs}
\label{S1:CorrGraph}

\begin{figure}
  \includegraphics[scale=0.25]{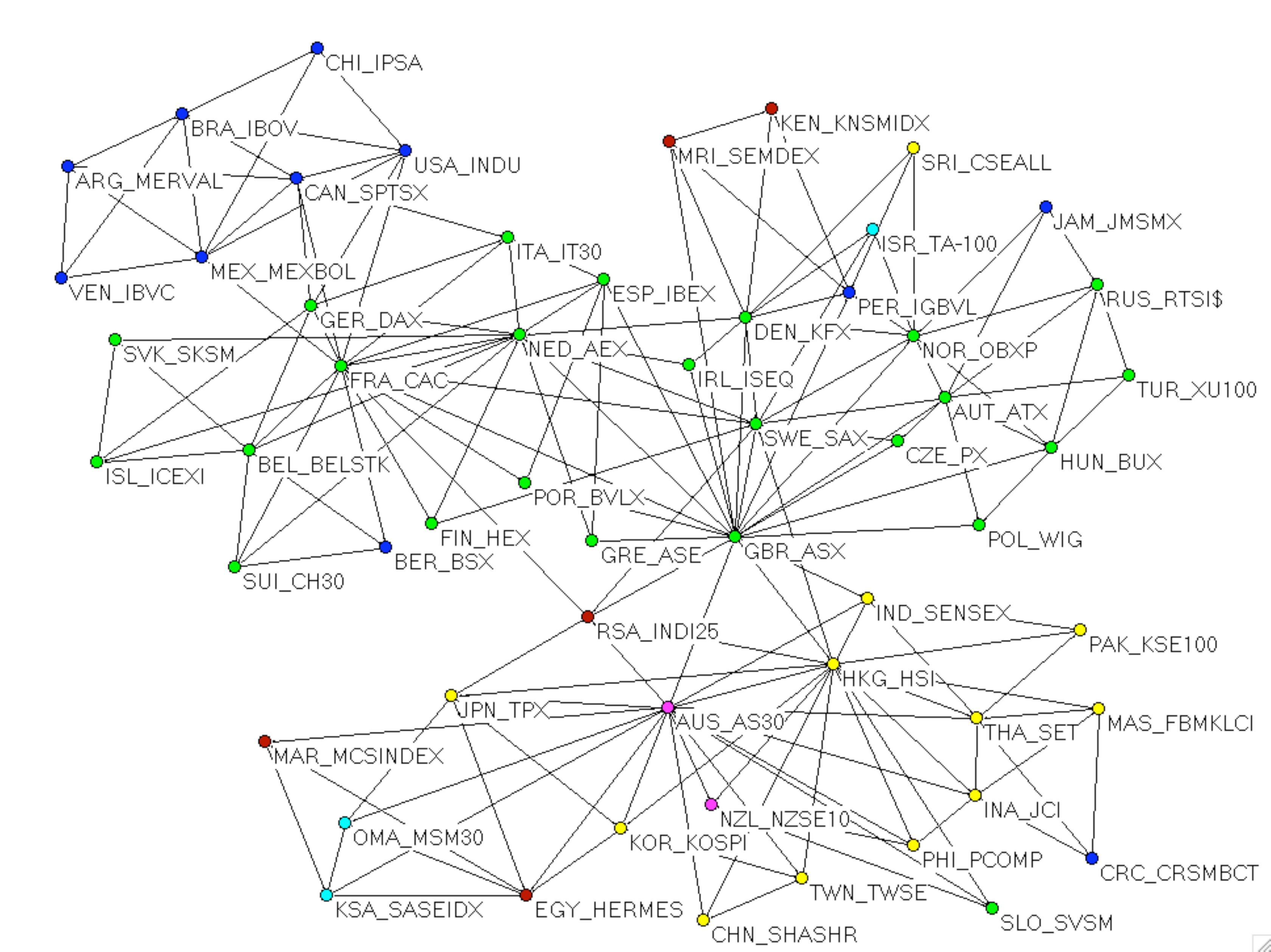}
  \caption{PMFG of the set of 57 market indices obtained from the correlation matrix of daily index returns estimated by using all daily records of the Jan 1996 - Jul 2009 time period. The color of symbols indicates the geographical location of the country hosting the stock exchange of the market index. American stock exchanges are blue, European green, Asian yellow, Oceanian magenta, Middle East cyan, and African maroon.}
  \label{figPMFG}
\end{figure}

Correlation based graphs provide a powerful tool to detect, analyze and visualize part of the most robust information which is present in the correlation matrix \cite{Tumminello-Lillo-Mantegna-2010-JEBO}. Here we first investigate the PMFG of the 57 selected market indices obtained from the correlation matrix estimated by using all the daily records of the selected time period (1/1/1996 - 31/7/2009). The unconditional PMFG is shown in Fig. \ref{figPMFG}. As already observed in previous studies \cite{Eryigit-Eryigit-2009-PA}, the relationship between market indices pointed out by the PMFG is primarily of geographical origin. On the top left of the graph we recognize the market indices of American stock exchanges (blue circles), market indices of European stock exchanges (green circles) are found in the central part of the graph and the bottom part of the graph links primarily market indices of Asian (yellow), Oceanian (magenta), middle east (cyan) and African (maroon) stock exchanges. An unsupervised cluster analysis of the indices can be performed on the PMFG by applying a community detection algorithm used to find community of elements in networks (for a recent review on this topic see Ref. \cite{Fortunato-2010-PR}).

Specifically, we investigate the community of elements of the PMFG by using the Infomap method proposed by Rosvall and Bergstrom \cite{Rosvall-Bergstrom-2008-PNAS}. This algorithm is considered one of the best algorithms of community detection in networks \cite{Lancichinetti-Fortunato-2009-PRE}. The method uses the probability flow of random walks  to identify the community structure of the system in the investigated network. This approach implies that two independent applications of the method to the same network may produce (typically slightly) different partitions of vertices. We repeat the application of the method 100 times to detect a minimum value of the fitness parameter estimating the goodness of the partition.

\begin{figure}
\includegraphics[scale=0.25]{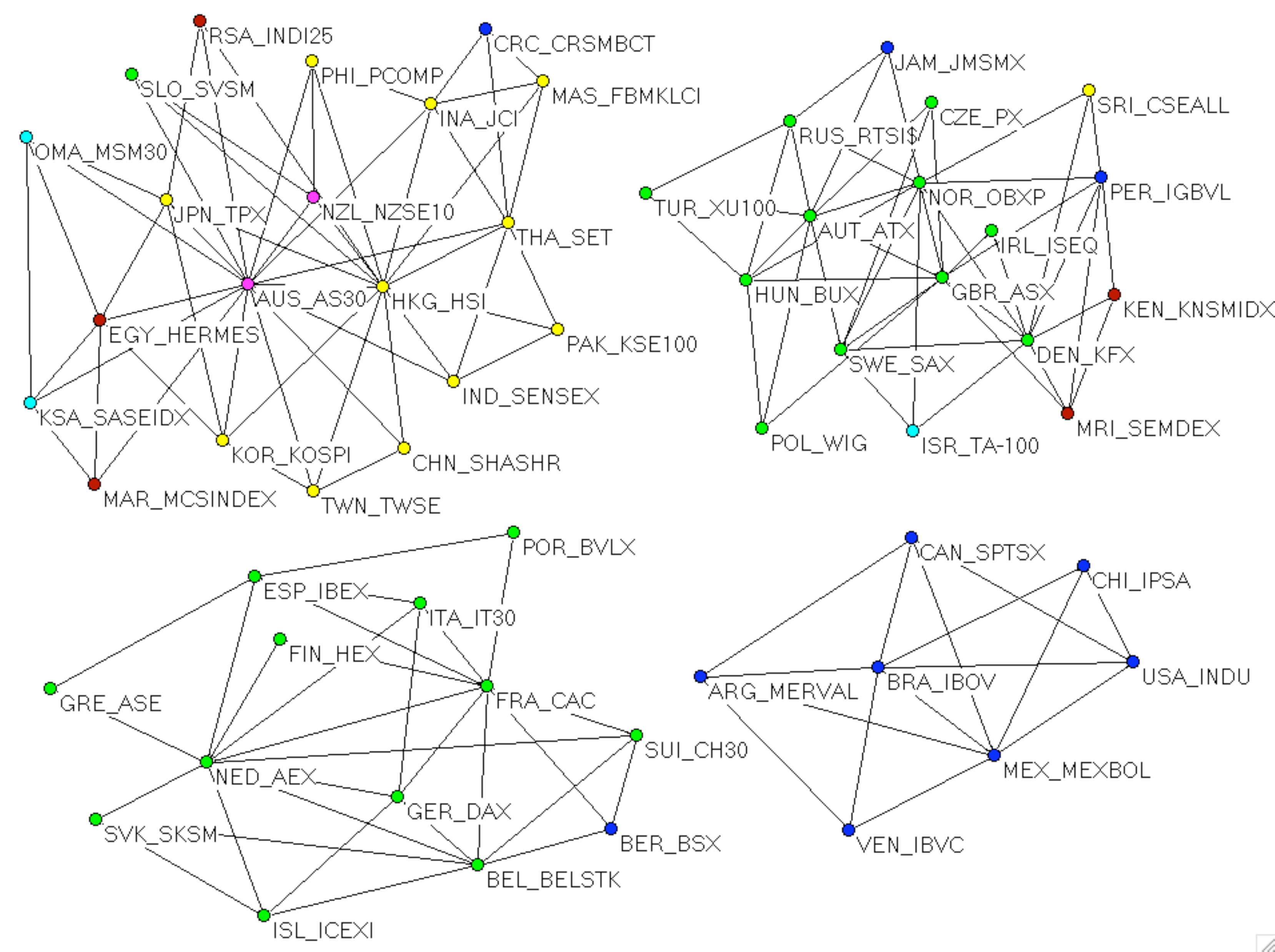}
\caption{Clusters (communities) of market indices detected into the unconditional PMFG by the Infomap community detection algorithm. Colors of the symbols are as in Fig. \ref{figPMFG}. The four detected clusters correspond primarily to different geographical regions. Top left: Asia and Oceania. Top right: North and East Europe. Bottom left: Central and Southern Europe. Bottom right: America.}
\label{figInfomap}
\end{figure}

The result of the application of the method to the unconditional PMFG is given in Fig. \ref{figInfomap}. The method identifies four distinct clusters. The bottom right cluster of the figure is the cluster of American stock exchanges. Two other clusters (top right and bottom left in the figure) are clusters of primarily European stock exchanges, whereas the forth cluster (top left in the figure) is primarily composed by Asian and Oceanian stock exchanges. In the following, we will use this unsupervised classification when we present the results of our analysis on the fast dynamics of correlations. In fact, the remaining part of this paper is devoted to an analysis of the properties of the correlation matrices and of the correlation based graphs estimated by using short evaluation time periods with a limited number of daily records so that they might provide information about the fast dynamics of correlation as a function of time.

\section{Short term dynamics of the correlation and of correlation based graphs}
\label{S1:Events}

\begin{figure}
\includegraphics[scale=0.09]{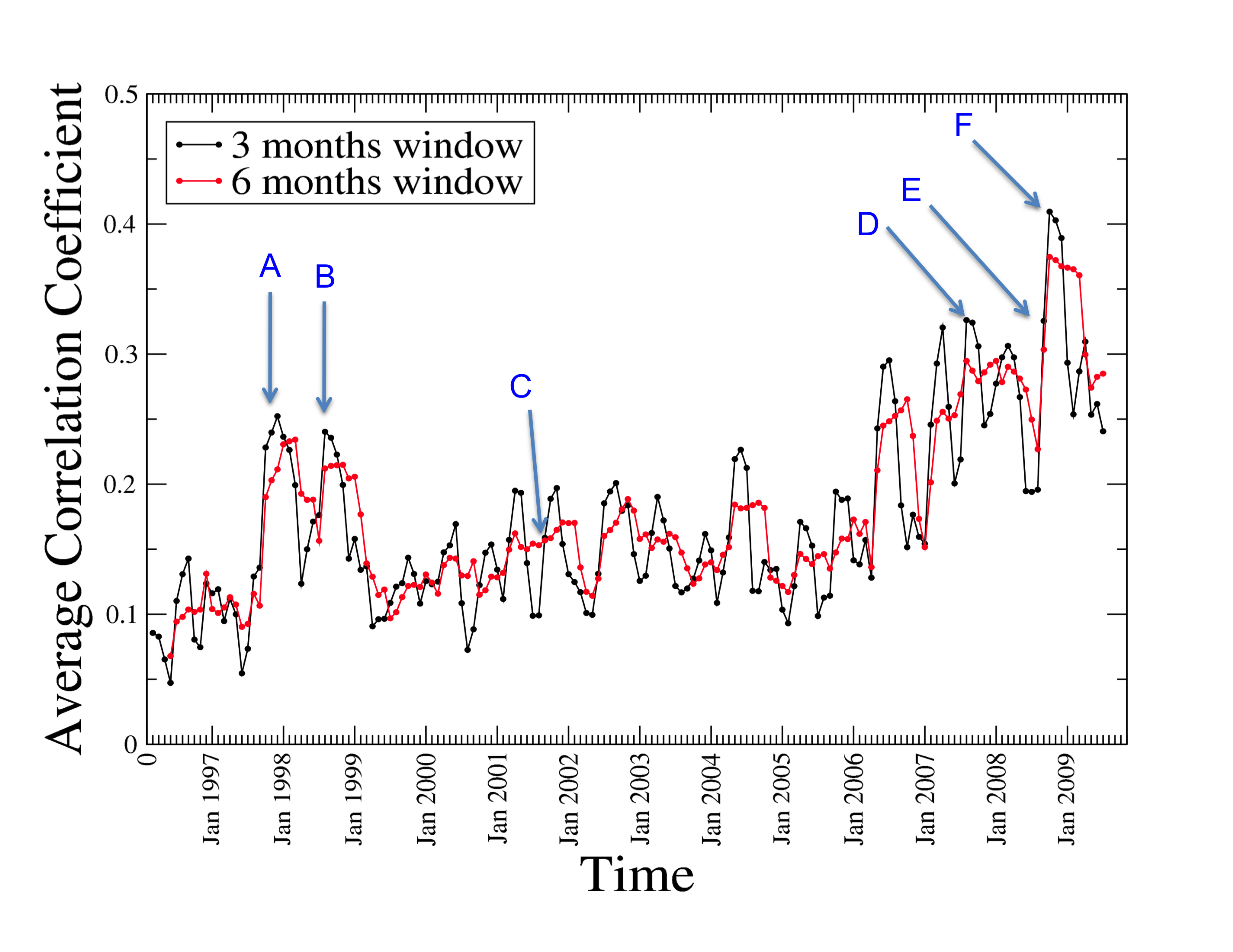}
\caption{Time evolution of the average correlation of the non diagonal elements of correlation matrices estimated by using an evaluation time period of 6 months (red line and symbols) or 3 months (black line and symbols). The fast dynamics is better described by the shortest evaluation time period. Labels A to F refer to the following events: A) Asian crisis 10/1997, B) Russian crisis 8/98, C) 9/11 terrorist attack 9/2001, D) onset subprime 8/2007, E) Lehman's failure 9/2008, F) onset of the global financial crisis 10/2008.}
\label{fig3m6m}
\end{figure}

In Fig. \ref{fig3m6m} we show the average correlation of the non diagonal elements of correlation matrices estimated for the 57 selected indices by using the evaluation time period of 6 months ($\Delta T=0.5$) and the shortest evaluation time period of 3 months ($\Delta T=0.25$) which is the shortest accessible for this set of indices by requiring that the correlation matrix is invertible (in fact a 3 month window typically presents in average 63.7 daily records a number which is only 1.12 times the number of indices of the investigated set).
The figure clearly shows that the time scale of the average correlation among indices is certainly shorter than six trading months. By using an evaluation time window of 6 months we already observe the smearing out of the correlation dynamics. A detailed analysis of some prominent financial crises clearly supports our conclusion. In fact in Fig. \ref{fig3m6m} the analysis of the 1997 Asian crisis (see arrow labeled as A in the figure) and of the 1998 Russian crisis (labeled as B in the figure) is quite resolved only when the 3 month evaluation time period is used. Similarly, the September 11, 2001 shock is visible as a sharp increase of the average correlation (labeled as C in the figure) only when the 3 month evaluation time period is used.  The onset of the subprime crisis (labeled as D in the figure), the Lehman's failure (labeled as E in the figure) and the peak of onset of the recent global financial crisis (labeled as F in the figure) are much more resolved again when the evaluation time period is 3 months.

We therefore conclude that a short time scale of less than 3 trading months is present in the time evolution of the dynamics of correlation coefficient of market indices of stock exchanges located all over the world. In the following Sections we will focus our attention on the changes of the correlation matrix estimated by using a 3 month evaluation time period.

\section{Dynamics of the PMFGs}
\label{S1:MutualInfo}

For each month of the investigated time period ranging from March 1996 to Jul 2009 we estimate a correlation matrix by using a 3 month interval comprising in average 63.7 daily records (for example the first record is computed by using the daily records of the period 1/1/1996 - 31/3/1996). From each correlation matrix we construct the PMFG and we investigate how links changes from month to month. Specifically, we consider the mutual information of links computed between two successive PMFGs. The way we compute the mutual information of links is explained in the following subsection.

\subsection{Mutual information}

We consider two networks with the same vertices, but in general with different sets of links. Let $N$ be the number of vertices in both networks. Let us indicate the number of links in the first network with $n_1$, and the number of links in the second network with $n_2$. We associate a binary random variable $x$ with all pair of vertices in the first network and a binary random variable $y$ with all pair of vertices in the second network. The variable $x$ takes the value 1 if two vertices are linked in the first network and it is 0
otherwise. Similarly $y$ describes links between vertices of the second network. The probability $p_1 (1)$ ($p_2(1)$) is the probability that a randomly selected pair of vertices is linked in the first (second) network. This definition implies that:
\begin{eqnarray*}
  p_1(1)&=&2n_1/(N^2-N) \\
  p_1(0)&=&1-p_1(1) \\
  p_2(1)&=&2n_2/(N^2-N) \\
  p_2(0)&=&1-p_2(1) 
\end{eqnarray*}
The joined probability $p(x,y)$ of the two variables $x$ and $y$ is given by:
\begin{eqnarray*}
  p(1,1)&=&2n_{1,2}/(N^2-N)\\
  p(1,0)&=&2(n_1-n_{1,2})/(N^2-N)\\
  p(0,1)&=&2(n_2-n_{1,2})/(N^2-N)\\
  p(0,0)&=&1-2(n_1+n_2-n_{1,2})/(N^2-N)
\end{eqnarray*}
where $n_{1,2}$ is the number of same links which are present in both networks. The mutual information of the random variables $x$ and $y$ is given by:
\begin{equation}
I(x,y)= \sum_{x=0,1} \sum_{y=0,1} p(x,y) \log \frac{p(x,y)}{p_1(x)p_2(y)}
\end{equation}
The mutual information $I(x,y)$ can be suitably normalized by dividing it for the geometric mean of the entropies $H(x)$ and $H(y)$ \cite{Strehl-2002,Yao-2003}:
\begin{equation}
  i(x,y)=I(x,y)/\sqrt{H(x)H(y)}
\end{equation}
where $H(x)$ is the entropy of variable $x$ and $H(y)$ is the entropy of variable $y$:
\begin{eqnarray*}
  H(x)=-p_1(0) \log p_1(0)-p_1(1) \log p_1 (1) \\
  H(y)=-p_2(0) \log p_2(0)-p_2(1) \log p_2 (1) \\
\end{eqnarray*}
It is to notice that the normalized mutual information $i(x,y) $ between identical networks is equal to 1.

\begin{figure}[b]
  \includegraphics[scale=0.35]{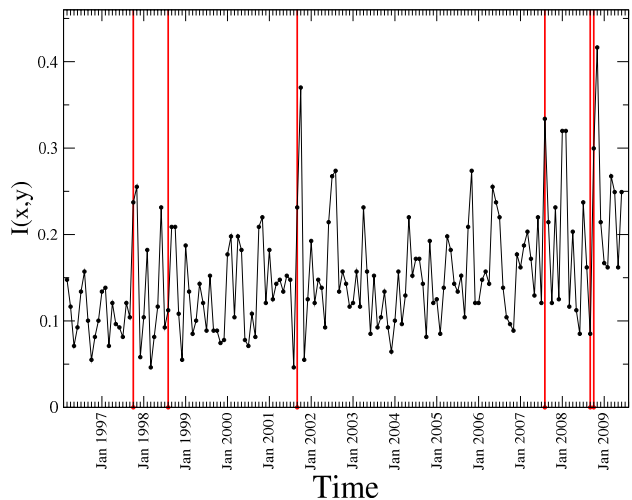}
  \caption{Mutual information of links between a PMFG estimated at month $t$ and the PMFG estimated at the successive month $t+1$. Vertical lines indicate events A to F highlighted in Fig. \ref{fig3m6m}.}
  \label{figMD}
\end{figure}

\subsection{Empirical analysis of PMFG graphs}

In Fig. \ref{figMD} we show the mutual information between the PMFG at month $t$ and the PMFG at the successive month $t+1$. In the figure we also highlight for reference the months when events A to F described in Fig.~\ref{fig3m6m} occurs. From the figure we notice that the structure of the PMFG is significantly altered during these months of big events. In fact, the correlation based graphs carry relevant information about the correlation profile of the indices. We now move to the analysis of (i) the time evolution of the degree profile of the different market indices in the PMFGs and (ii) the assessment of the statistical differences observed between the set of links defined by different correlation based graphs.

\begin{figure}[t]
\includegraphics[scale=0.07]{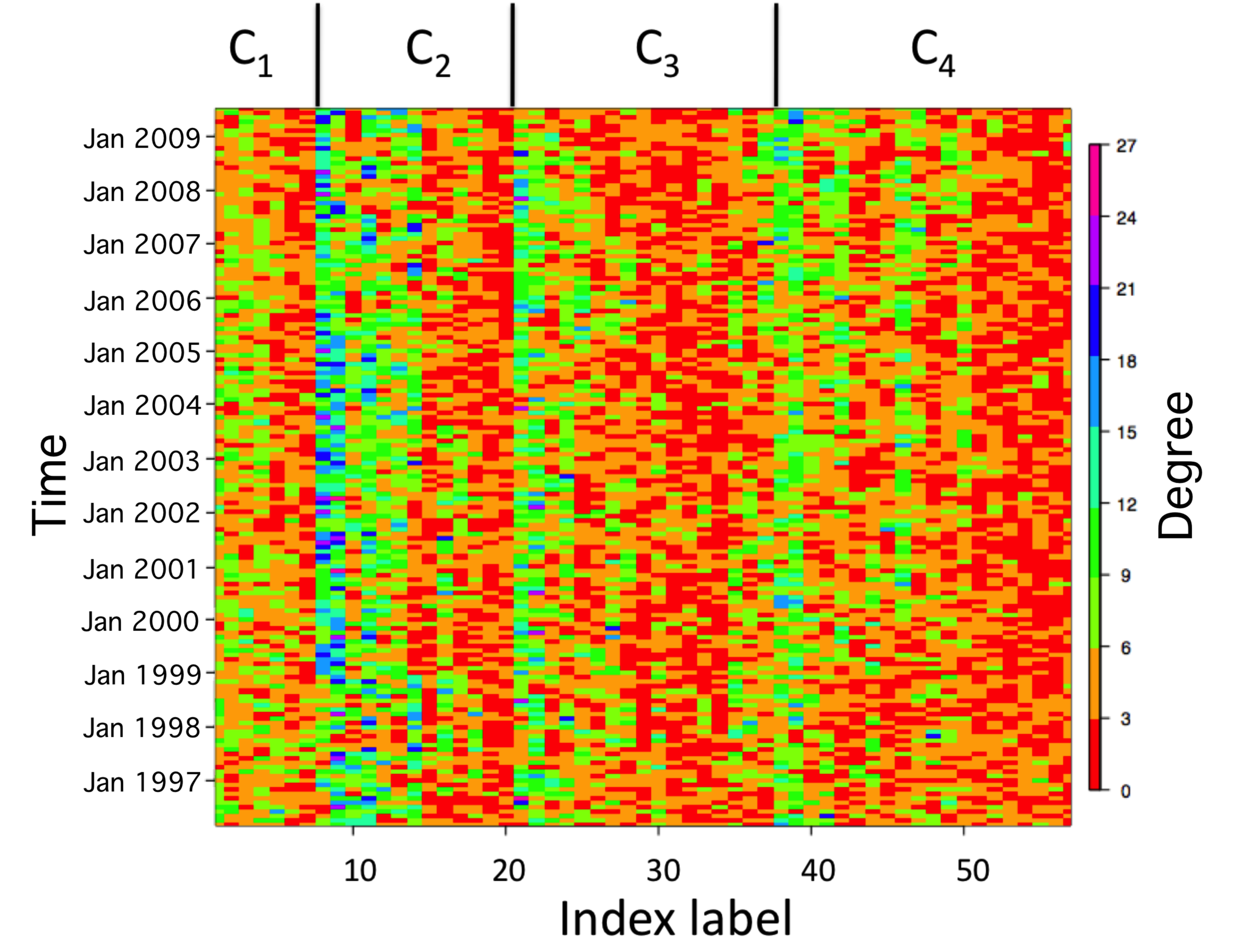}
\caption{Color code representation of the time evolution of the degree of market indices of the PMFGs computed by using a 3 months evaluation time period. Market indices are ordered from left to right according to the clusters detected by the Infomap algorithm in the unconditional PMFGs of Fig. \ref{figPMFG}. $C_1$ to $C_4$ clusters are the clusters shown in Fig. \ref{figInfomap}.}
\label{figDTE}
\end{figure}

The result of the first investigation is summarized in Fig. \ref{figDTE}. In the figure we show a color code representation of the time evolution of the degree of each market index observed in the PMFGs computed for all the 161 investigated months. Different market indices are ordered accordingly to the rank of the four clusters obtained by the Infomap partitioning of the unconditional PMFG computed by using the correlation matrix estimated using all daily records (see Fig. \ref{figPMFG} and Fig. \ref{figInfomap} of Section \ref{S1:CorrGraph}). Specifically, cluster 1 ($C_1$) is the cluster of American market indices, cluster 2 ($C_2$) is  a cluster of primarily European indices with market indices from continental and Mediterranean countries, cluster 3 ($C_3$) is  a cluster of primarily European indices with market indices from UK, Ireland and continental and North European countries, and cluster 4 ($C_4$) is a cluster of Oceanian and Asian market indices. By analyzing Fig. \ref{figDTE} we notice an overall persistence of the level of degree specific to each index. In particular the indices of highest rank within each cluster (rank is given within each cluster by the Infomap algorithm and reflects the role played by the element in the cluster detection), which are located at the left side of each cluster area are characterized by higher degree. Examples are market indices of  France (label 8), Netherlands (label 9), Germany (label 11) and UK (label 21) in Europe and indices of Australia (label 38) and Hong Kong (label 39) in the Pacific-Asian region.

\begin{figure}
\includegraphics[scale=0.35]{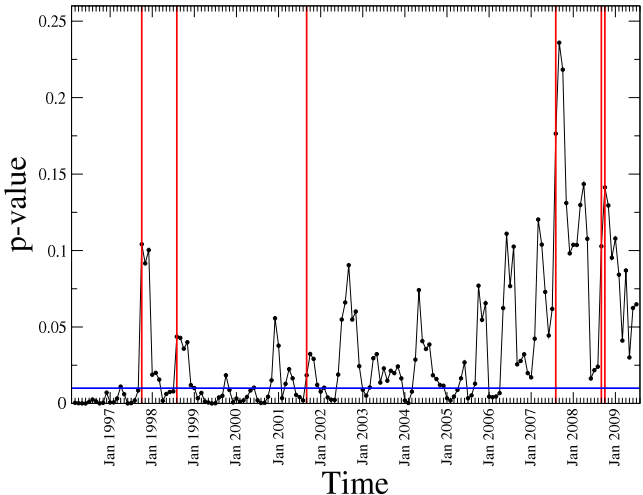}
\caption{P-value of Welch's t-test comparing the average values of correlation coefficients of MST's links and of PMFG's links over time. The vertical lines indicate events A to F of Fig. \ref{fig3m6m} whereas the horizontal line indicates a 0.01 threshold.}
\label{figpvalue}
\end{figure}

The second of our investigations shows that an alteration of the structure of the correlation based graphs is present around specific months. Specifically, a t-test for difference in mean is used to compare the correlation values associated with the links of two basic correlation based graphs, namely the PMFG so far discussed and the Minimum Spanning Tree (MST) \cite{Mantegna-1999-EPJB}. The p-value provided by the test is reported for each month of the investigated period in Fig. \ref{figpvalue}. The p-value is larger than 5\% for all the considered 5 crises, indicating that the average correlations of PMFG's links and MST's links are statistically consistent. A different behavior is observed in those periods of time not characterized by widely spread financial crises.
We interpret this second result as a manifestation of a significant alteration of the overall structure of the correlation matrix. The nature of these changes in the PMFG structure seems not to be of topological nature (in fact the degree profile of Fig. \ref{figDTE} is quite stable during time evolution) but rather might involve specific links. We have not been able so far to interpret in a simple and convincing way these changes, mainly due to the high level of statistical uncertainty
associated with the need of a short evaluation time period. In other words, we are able to see that useful information is there but it is dressed with a relevant level of noise unavoidably reflecting the statistical uncertainty associated with the correlation matrix estimation. Specific alterations associated with specific crises cannot be reliably detected without a procedure assessing the statistical robustness of each link.

\section{Spectral analysis of the correlation matrices}
\label{S1:Spectral}

\begin{figure}
\includegraphics[scale=0.35]{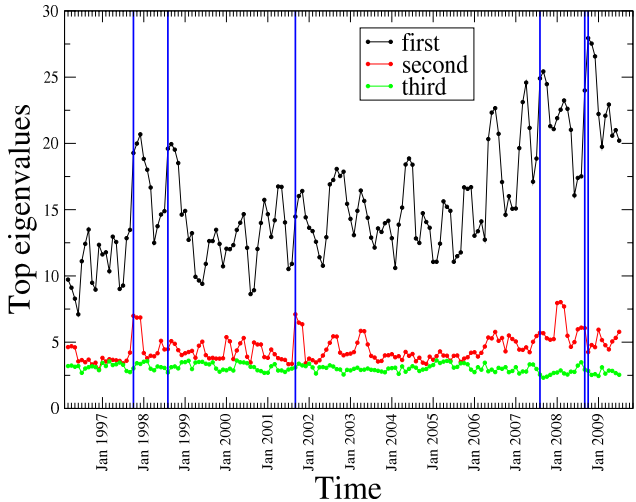}
\caption{Time evolution of the first, second and third eigenvalue of correlation matrices computed by using a 3 months evaluation time period. The vertical lines indicates events A to F of Fig.~\ref{fig3m6m}.}
\label{figTE}
\end{figure}

We lastly complement our analysis of the correlation based graphs with a spectral analysis of the correlation matrices. In our analysis we mainly focus on the time dynamics of the largest eigenvalues and of their corresponding eigenvectors. In Fig. \ref{figTE} we show the time evolution  of the first, second and third eigenvalues of the correlation matrices computed with a 3 months evaluation time period. The time profile of the first eigenvalue is highly correlated with time profile of the average correlation. The second eigenvalue shows abrupt changes in the presence of, or immediately after in the case of event C, special events (A-F) such as the ones highlighted in the figure. The third eigenvalue has a more limited excursion and it is unclear whether it carries information. In fact, the average number of eigenvalues above the random matrix theory threshold determined as suggested in Ref. \cite{Laloux-Cizean-Bouchaud-Potters-1999-PRL,Plerou-Gopikrishnan-Rosenow-Amaral-Stanley-1999-PRL} is equal to 3.00 and its standard deviation is 0.65. Therefore the first two eigenvalues are the only large eigenvalues whose presence cannot be consistent with a statistical uncertainty of the correlation matrix due to the finiteness of the market index time series. Again we conclude that information not compatible with a random null hypothesis is therefore present in these correlation matrices in spite of the high degree of statistical uncertainty associated with their estimation.

\begin{figure}
\includegraphics[scale=0.07]{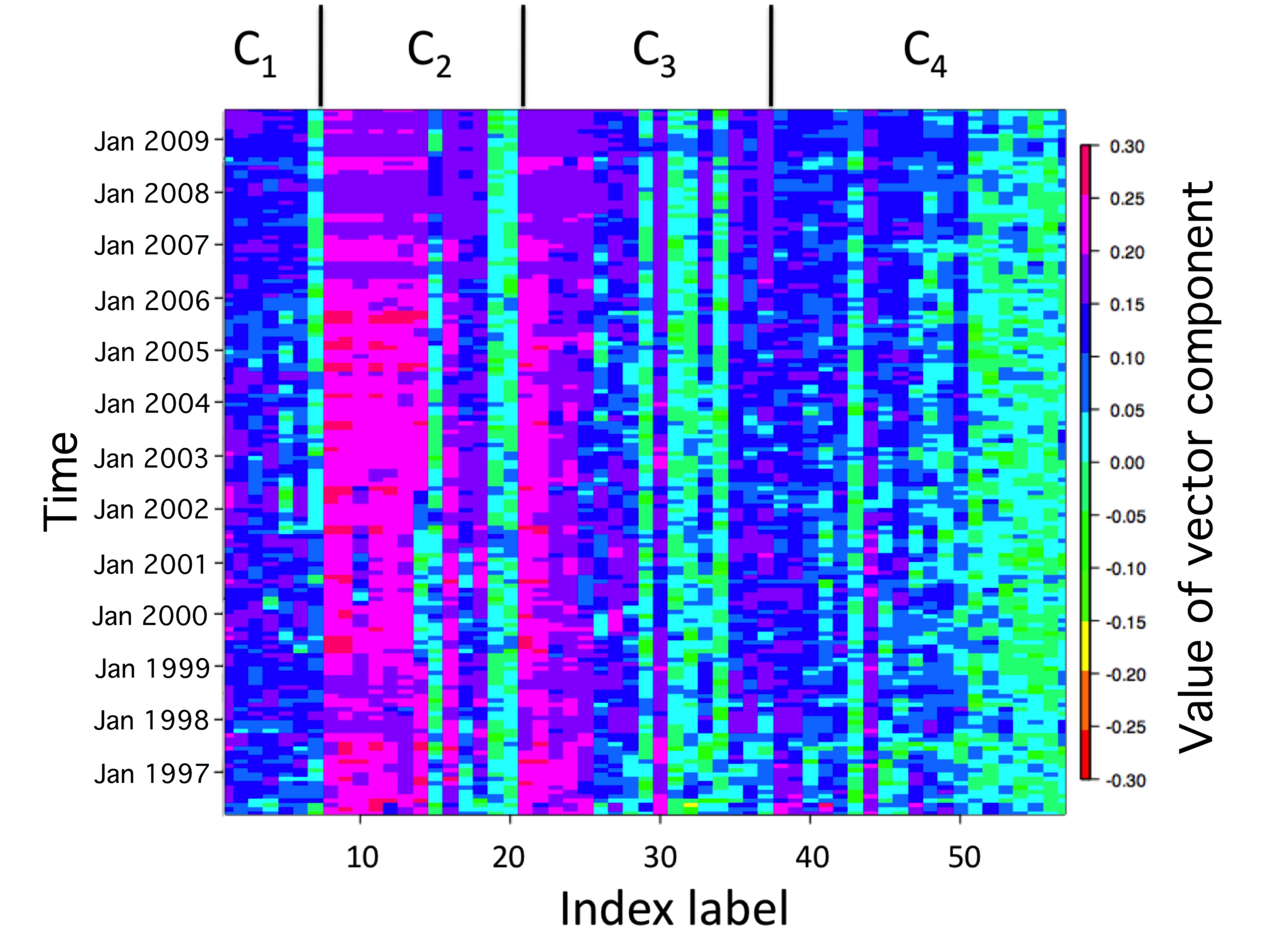}
\caption{Color code representation of the components of the first eigenvector as a function of time (vertical axis). The direction of the eigenvector is selected by making positive the component of the USA market index. Market indices are ordered from left to right according to the clusters detected by the Infomap algorithm in the unconditional PMFGs of Fig.~\ref{figPMFG}. $C_1$ to $C_4$ are the clusters shown in Fig.~\ref{figInfomap}.}
\label{figFE}
\end{figure}

One way to investigate the nature of this information is to analyze the profile of the eigenvectors associated with the largest eigenvalues. In Fig.~\ref{figFE} we show a color code representation of the components of the eigenvector associated with the first eigenvalue for all the 161 investigated months. The direction of the eigenvector is arbitrary. In the figure we select the direction associated with a positive component of the USA market index as the positive direction. The eigenvector components are mainly positive indicating the presence of a common factor driving a large number of market indices. This driving factor has high positive components in the majority of the European indices. American and Asian indices also show medium to high positive components. Negligible components or negative components are observed for some indices of emergent countries located in Europe, Middle East, Africa and Asia. In summary the components of the first eigenvector reflect a common factor driving mature markets located in all continents.

\begin{figure}
\includegraphics[scale=0.07]{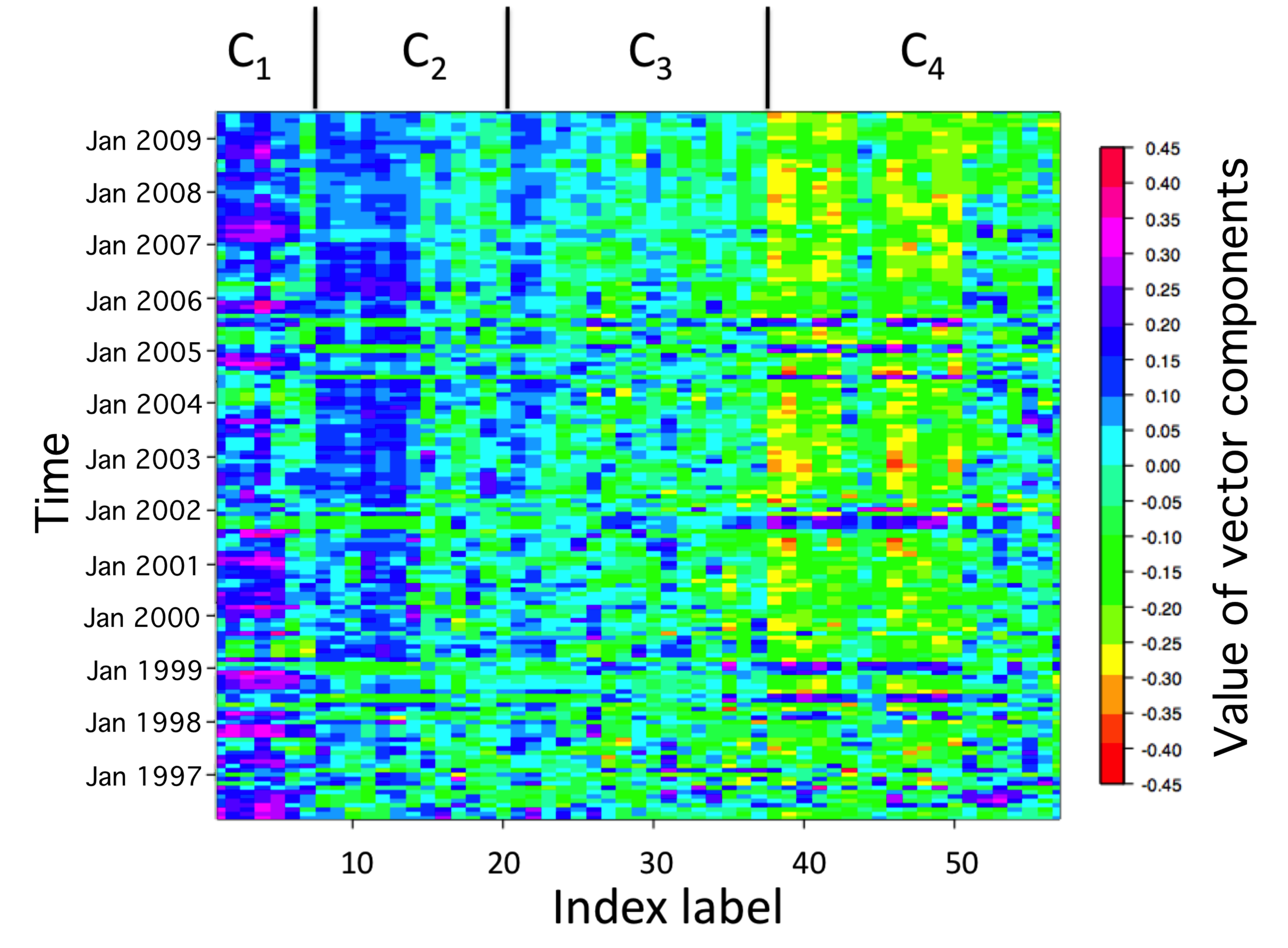}
\caption{Color code representation of the components of the second eigenvector as a function of time (vertical axis). The direction of the eigenvector is selected by making positive the component of the USA market index. Market indices are ordered from left to right according to the clusters detected by the Infomap algorithm in the unconditional PMFGs of Fig. \ref{figPMFG}. $C_1$ to $C_4$ are the clusters shown in Fig. \ref{figInfomap}.}
\label{figSE}
\end{figure}

Similarly to the case of the first eigenvector, in Fig.~\ref{figSE} we show a color code representation of the components of the second eigenvector. Also in this case, the positive direction of the eigenvector is associated with a positive component of the USA index. The components of the second eigenvector have a more complex structure than the ones of the first eigenvector. In fact we note that Asian and Oceanian market indices have components characterized by a sign opposite to the sign of American and some European indices. In other words the factor associated with this second eigenvalue is affecting indices of different regions of the world in a different manner. Differences are more pronounced between Asia-Oceania and Europe-America but also differences between European and American indices are sometimes observed. The behavior of the European indices is not as homogeneous as it is in the case of the components of the first eigenvector.

In summary, our analysis of the first two largest eigenvalues and eigenvectors of the correlation matrices shows that relevant information is present in them and in their dynamics. Two global factors are present, the first affecting primarily mature markets and the second discriminating quite well between market indices of different world regions.

\section{Conclusions}
\label{S1:Conclusion}

In this paper we investigate the daily correlation present among market indices of stock exchanges located all over the world. The study is performed by using the index time series of 57 different stock exchanges located all over the world and continuously monitored during the time period Jan 1996 - Jul 2009. By investigating this set of market indices we discover that the correlation among market indices presents both a fast and a slow dynamics. The slow dynamics is a gradual growth associated with the development and consolidation of globalization. We show that the fast dynamics is associated with events that originate in a specific part of the world and rapidly affect the global system. We provide evidence that the short term timescale of correlation among market indices is quite fast and less than 3 trading months (about 60 trading days). By computing correlation matrices each trading month using a 3 months evaluation time period we show that correlation matrices contain information about the world global system that can be investigated by using average values of the correlation, correlation based graphs and the spectral properties of the largest eigenvalues and eigenvectors. The overall changes of the correlation based graphs are investigated by using a newly introduced mutual information of link co-occurrence in networks with the same number of elements. Changes affecting specific links during prominent crises are of difficult interpretation due to the high level of statistical uncertainty associated with the correlation estimation and because successive rewiring of links  might be crisis specific and therefore specific to each single event. In a future study we aim to achieve a more robust statistical validation of the rewiring of links occurring in the presence of short term abrupt changes of correlation profile with method based on bootstrap \cite{Tumminello-Coronnello-Lillo-Micciche-Mantegna-2007-IJBC}.

\begin{acknowledgments}
We thank Ken Bastiaensen for providing the stock index data. DMS and WXZ acknowledge financial support from the National Natural Science Foundation of China (11075054) and the Fundamental Research Funds for the Central Universities. RNM acknowledges financial support from the PRIN project 2007TKLTSR ``Indagine di fatti stilizzati e delle strategie risultanti di agenti e istituzioni osservate in mercati finanziari reali ed artificiali''.
\end{acknowledgments}

\appendix*\section{Set of market indices}

We investigate the daily synchronous dynamics of 57 stock market indices located in 57 different countries. The countries and stock indices investigated are: Argentina (MERVAL), Australia	(AS30), Austria (ATX), Belgium (BEL20), Bermuda (BSX?), Brazil (IBOV), Canada (SPTSX), Chile (IPSA), China (SHASHR), Costa Rica (CRSMBCT), Czech Republic	(PX), Denmark (OMX COPENHAGEN 20), Egypt (HERMES), Spain (IBEX 35), Finland (OMX HELSINKI), France	(CAC 40), Germany (DAX), Greece (Athex Composite), Hong Kong (HANG SENG), Hungary (BUX), Indonesia (JAKARTA COMPOSITE), India (SENSEX 30), Ireland (ISEQ), Iceland	(OMX Iceland All-Share), Israel (TA-100), Italy (IT30), Jamaica (JMSMX), Japan (TPX), Kenya (KNSMIDX), Korea (KOSPI), Saudi Arabia (SASEIDX), Morocco (CFG25), Malaysia (FTSE Bursa Malaysia), Mexico (IPC), Mauritius (SEMDEX), Netherlands (AEX), Norway (OBX), New Zealand (NZSE10), Oman (MSM30), Pakistan (KSE100), Peru (IGBVL), Philippines (PSEi), Poland (WIG), Portugal (PSI GENERAL), South Africa (INDI25), Russia (RTSI), Slovenia (SBI20), Sri Lanka (CSEALL), Switzerland (CH30), Slovakia (SKSM), Sweden (OMX STOCKHOLM), Thailand (SET), Turkey (XU100), Taiwan	(TAIEX), UK (FTSE ALL-SHARE), United States (DOW JONES INDUS.), Venezuela (IBVC).

\end{document}